\documentclass[12pt]{article}
\usepackage[utf8]{inputenc}
\usepackage[margin=1in]{geometry}
\usepackage{amsfonts}
\usepackage{natbib}
\usepackage{bm}
\usepackage{amsmath}
\usepackage{amssymb}
\usepackage{amsthm}
\usepackage{mathtools}
\usepackage{url}
\usepackage{cancel}
\usepackage{eurosym}
\usepackage{float}

\usepackage{bbm}
\usepackage{authblk}
\usepackage{color}
\usepackage{xcolor,colortbl}
\allowdisplaybreaks

\title{A Bayesian spatial hierarchical model for extreme precipitation in Great Britain}

\author{Paul Sharkey$^{1}$ and Hugo C. Winter$^{2}$ \\
Email: \texttt{p.sharkey1@lancaster.ac.uk}; \texttt{hugo.winter@edfenergy.com}}
\affil{$^{1}$\textit{STOR-i Centre for Doctoral Training, Department of Mathematics and Statistics, Lancaster University, Lancaster, UK.} \\ $^{2}$\textit{EDF Energy R\&D UK Centre, Interchange, 81-85 Station Road, Croydon, CR0 2AJ, UK.}}
\date{}

\begin{document}

\maketitle

\begin{abstract}
Intense precipitation events are commonly known to be associated with an increased risk of flooding. As a result of the societal and infrastructural risks linked with flooding, extremes of precipitation require careful modelling.  Extreme value analysis is typically used to model large precipitation events, though a site-by-site analysis tends to produce spatially inconsistent risk estimates. In reality, one would expect neighbouring locations to have more similar risk estimates than locations separated by large distances. We present an approach, in the Bayesian hierarchical modelling framework, that imposes a spatial structure on the parameters of a generalised Pareto distribution. In particular, we look at the clear benefits of this approach in improving spatial consistency of return level estimates and increasing precision of these estimates. Unlike many previous approaches that pool information over locations, we account for the spatial dependence of the data in our confidence intervals. We implement this model for gridded precipitation data over Great Britain.     
\end{abstract}
\hspace{3.5mm}{\bf Keywords:} Extreme value analysis, spatial modelling, covariate modelling, climate extremes, Bayesian inference.

\section{Introduction}
\label{sec:intro}
In a changing climate with an increased frequency of intense precipitation events \citep{trenberth2011changes}, modelling the rate and size of such events has become increasingly important. In Great Britain, extreme events can arise from the presence of extratropical cyclones evolving from the North Atlantic Ocean. They can also originate from short-term localised convective behaviour. Extreme precipitation levels are commonly associated with an increased risk of flooding \citep{kunkel1999long}, which can contribute to substantial infrastructural damage. Consequently, estimation of extreme precipitation is a vital component of hydrological models for assessing flood risk and therefore needs to be modelled carefully.  \\

Extreme value analysis is used in practice to model rare events by extrapolating beyond observed data to give probability estimates of events occurring at unobserved levels. In this way, one can make predictions of future extreme behaviour by estimating
the behaviour of the process using an asymptotically justified limit model. Let $X_1, \hdots, X_n$ be a sequence of independently and identically distributed (iid) random variables with distribution function $F$. Defining $M_n = \max(X_1,\hdots, X_n)$, if there exists sequences of constants $a_n > 0$ and $b_n$, such that, as $n \rightarrow \infty$
\begin{equation*}
    \Pr\left(\frac{M_n-b_n}{a_n} \leq x \right) \rightarrow G(x),
\end{equation*}
for some non-degenerate distribution $G$, then $G$ is a generalised extreme value (GEV) distribution with distribution function
\begin{equation*}
 G(x) = \exp \left\{- { \left[1+\xi \left( \frac{x-\mu}{\sigma} \right) \right] }_{+}^{-1/\xi} \right\},
 \label{eq:gev}
\end{equation*}
where $x_{+} = \max(x,0)$, $\sigma >0$ and $\mu, \xi \in \mathbb{R}$. In this formulation, $\mu, \sigma$ and $\xi$ denote location, scale and shape parameters respectively. \\

An alternative to modelling the maxima of random variables is to model excesses above a high threshold. Conditional on a high threshold $u$, the distribution of excesses above $u$ can be approximated by a generalised Pareto distribution \citep{pickands1975statistical} such that
\begin{equation} 
\Pr(X-u > x | X > u) = {\left(1+\frac{\xi x}{\sigma_{u}}\right)}^{-1/\xi}_{+}, \mbox{                                } x > 0,
\label{eq:gpd}
\end{equation}
where $\sigma_u>0$ denotes the threshold-dependent scale parameter and $\xi$ denotes the shape parameter, identical to that of the GEV distribution. A third parameter $\lambda_u$, denoting the rate of exceedance, must also be estimated. In practice, this approach to inference is often preferred to analysing block maxima as parameter uncertainty is reduced by utilising more extreme data. The threshold $u$ is typically chosen using standard diagnostics outlined in \citet{coles2001introduction}. An alternative, but equivalent approach for modelling threshold excesses is the Poisson process model \citep{coles2001introduction}, which is theoretically more suited to modelling extremes in the presence of covariates, but has more issues in implementation \citep{sharkeypoisson}. This work will use the GPD as our asymptotically justified threshold excess model.  \\

Extreme value models are typically constructed on a site-by-site basis, with no inherent spatial structure built into models and instead rely on the data to reveal any spatial similarity in the marginal distributions. Precipitation tends to affect clusters of sites simultaneously as a result of a generating storm system, for example. Intuitively, after accounting for physical effects relating to geography, one would expect the probability of extreme precipitation events to be more similar for neighbouring sites than for sites separated by large distances. Analysis of individual sites in isolation tends to produce spatially inconsistent probability estimates, which justifies the need for an extreme value model that incorporates spatial information. \\

A natural class of models for spatial extremes are max-stable processes, which are the extension of univariate and multivariate extreme value theory to the infinite-dimensional setting. In particular, the limiting process of the componentwise maxima of a sequence of normalised stochastic processes is a max-stable process. These are commonly used for modelling spatial extremes, where the underlying marginals are distributed as unit Fr\'echet. In practice, max-stable processes are difficult to fit due to the number of terms required in the likelihood computation. Max-stable processes are unsuitable for modelling variables that are independent in their extremes as they assume extremal dependence \citep{kereszturi2016assessing}. In addition, they are limited in the sense that they are only suitable for observations that are componentwise (e.g. annual) maxima. Recently, however, generalised Pareto processes have been used to extend the concept of threshold excess models to the space of continuous functions \citep{ferreira2014generalized}. For more details on max-stable processes, see \citet{smith1990max}, \citet{schlather2002models} and \citet{padoan2010likelihood}. \\

Another method widely used in the hydrology community is regional frequency analysis (RFA). The aim of RFA is to borrow strength across neighbouring locations with homogeneous statistical behaviour to reduce uncertainty in parameter estimates. It has been extensively studied in \citet{hosking2005regional}. It is a useful approach in extreme value analysis as the pooled information can increase confidence in return level estimates, which can often be highly uncertain due to the scarce nature of the data used in the analysis. The specification of homogeneous regions can be somewhat restrictive as it imposes artificial spatial boundaries on the quantity to be estimated, creating spatially inconsistent estimates across regions. A further disadvantage is that covariates cannot be implemented as part of the L-moments scheme used for estimation, meaning that physical information cannot be incorporated into the model in this manner. \\ 


Spatial Bayesian hierarchical models have been used in extreme value analysis with a similar aim to RFA - to use information from neighbouring locations to produce spatially consistent return level estimates and reduce uncertainty in these estimates. These methods are often used to model spatial count and binary data \citep{diggle1998model}. For a comprehensive overview of such methods, see \citet{banerjee2014hierarchical}. In recent years, these methods have been extended for use in extreme value analysis. \citet{cooley2007bayesian} modelled threshold excesses at 56 sites using the GPD with an underlying latent spatial model for the GPD parameters. This model was used to interpolate the GPD parameters over the entire domain. \citet{sang2009hierarchical} built a hierarchical model for extreme precipitation on a lattice, using an Intrinsic Autoregressive Model (IAR) as the latent process, but assumed that the shape parameter was constant over the region they studied. In this work, we use the model of \citet{cooley2010spatial}, which builds on the model of \citet{sang2009hierarchical} by modelling the shape parameter using a latent spatial process. \\

For extreme value analysis, Bayesian hierarchical models are advantageous in their flexibility and incorporation of physical and spatial information through covariates and random effects respectively. Because small proportions of data records are used in extreme value problems, the reduction in uncertainty gained from pooling over space is particularly useful. The Bayesian hierarchical framework relies on the assumption that the extremes are independent conditional on the covariate structure and latent process. While this makes the model unsuitable for modelling joint extremes, our interest lies in how the extremal characteristics of the marginal distribution of precipitation varies across locations, and we can compute return levels to that effect. We construct our likelihood function under the assumption of conditional independence with an adjustment factor \citep{ribatet2012bayesian}, which allows for appropriate inference and uncertainty quantification under this misspecified likelihood. Previous studies that pool information across sites, in both RFA and the hierarchical modelling framework, typically find unrealistically narrow estimated return level confidence intervals. We, however, account for the spatial dependence in the data in our uncertainty quantification and achieve more feasible confidence intervals.  \\

In this paper, we describe a spatial extreme value model using the Bayesian hierachical modelling framework, using an adjusted likelihood to account for the misspecification of conditional independence. By imposing a condition of spatial similarity on the model parameters, we can produce a spatially consistent map of probabilities of extreme events. Section~\ref{sec:data} details the precipitation data along with a previous study using the RFA approach. Section~\ref{sec:model} gives a comprehensive overview of our modelling strategy. In Section~\ref{sec:results}, we apply the hierarchical model to the data and compare the results with other approaches. We conclude in Section~\ref{sec:disc} with some discussion.

\section{Data}
\label{sec:data}

\subsection{Data description}
\label{subsec:datadesc}

\begin{figure}[h!]
    \centering
    \includegraphics[width=7cm,angle=90]{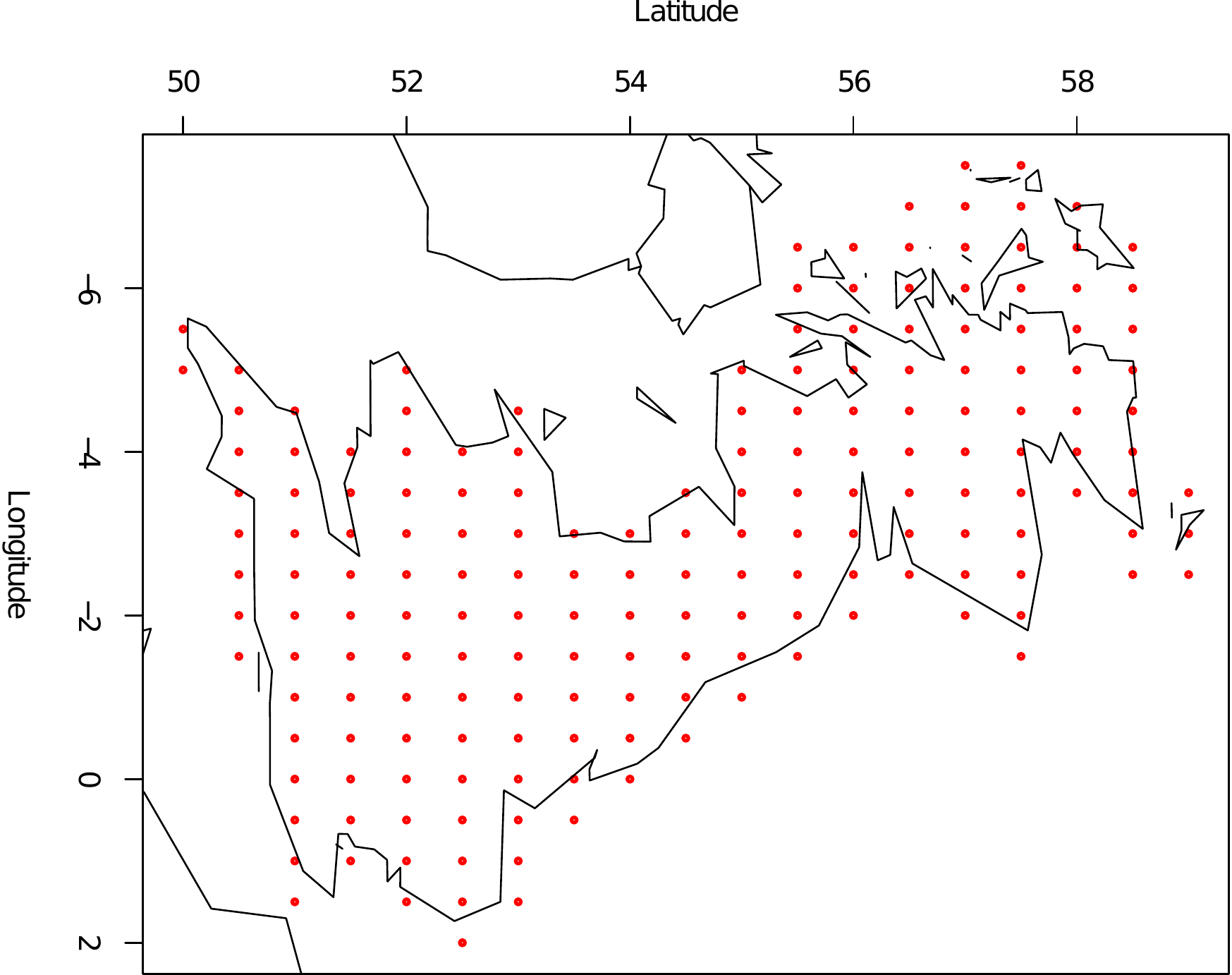}
    \caption{Locations of gridded reanalysis precipitation data used in the analysis}
    \label{fig:domain}
\end{figure}

Reanalysis precipitation data were taken from the Climate Forecast System Reanalysis dataset (CFSR), supplied by the National Centres for Environmental Prediction (NCEP). Input models for the reanalysis include various atmospheric, ocean and land models.
Data are available on a daily scale from January 1979 - August 2016 on a $0.5^{\circ}$ resolution grid (see Figure~\ref{fig:domain}). The data are spatially and temporally complete, thus does not contain missing values; this avoids any issues resulting from the treatment of missing data, however, there are limitations. Due to complex processing and the combination of data sources if it often difficult to ascertain the source of error, making uncertainty in the data difficult to quantify. \\

\begin{figure}[h!]
    \centering
    \includegraphics[width=10cm]{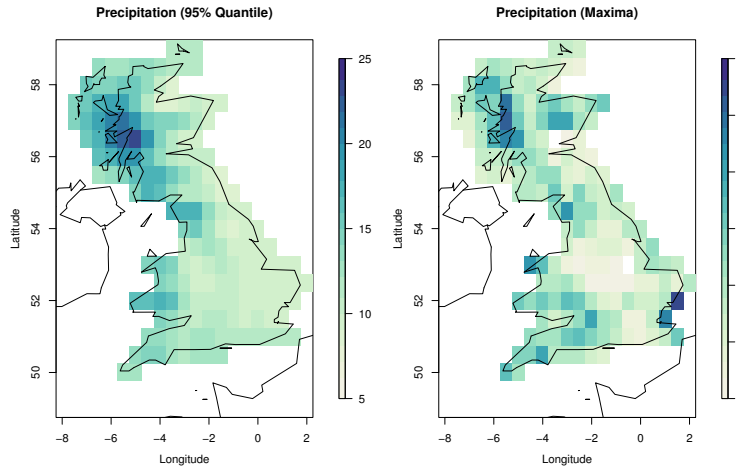}
    \caption{The 95$\%$ quantile (left) and maximum (right) of the precipitation data in each cell in 37 years of data.}
    \label{fig:exploratory}
\end{figure}

 Figure~\ref{fig:exploratory} shows that the largest precipitation events tend to occur in north-west Scotland, with a general tendency for larger events on the west coast of Great Britain, which one may expect due to the west-east passage of weather systems over the Atlantic Ocean and Irish Sea. The spatial distribution of maxima is more varied. The global maximum value occurs in south-east England, in a region that is typically dry, at least compared to Scotland. This event, which occurred during summer, was most likely a result of a short-term, localised convective event, whereas the precipitation in Scotland is typically the result of large-scale synoptic storms arising from the Atlantic Ocean.

\subsection{Regional frequency analysis}
A study of this dataset is conducted in \citet{winter2017rfa} using a regional frequency analysis (RFA) approach. This aims to reduce uncertainty in return level esimation by pooling information across regions with statistically homogeneous behaviour. The first step in this approach is to define such regions using techniques in \citet{weiss2014formation}. Observations are standardised by a cell-specific threshold and a regional GPD distribution~(\ref{eq:gpd}) is fitted. The cell-specific distribution is obtained by rescaling the regional model using the cell-specific threshold. Inference is carried out using L-moments estimation. \\

\begin{figure}[H]
    \centering
    \includegraphics[width=15cm]{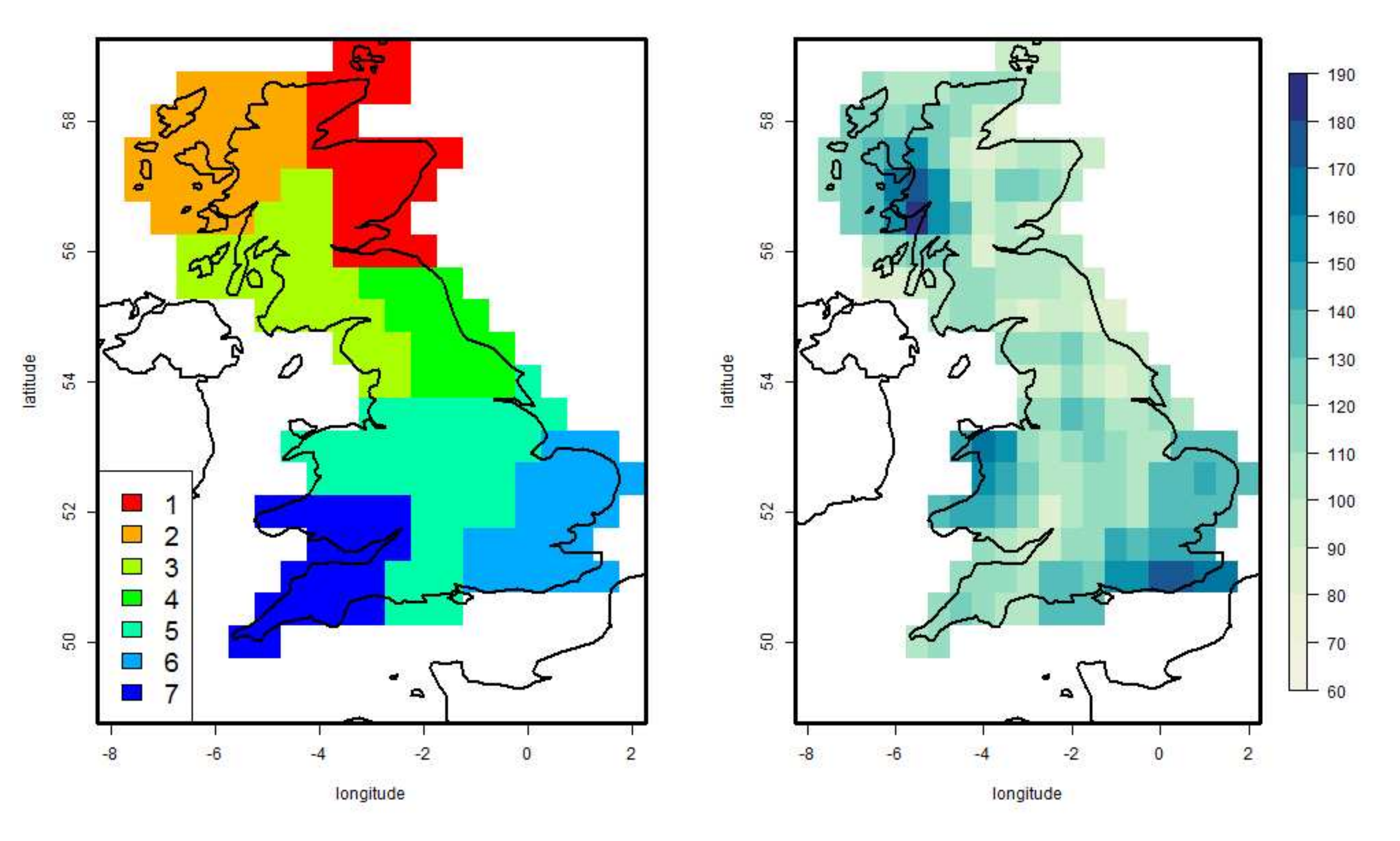}
    \caption{The homogeneous regions specified for RFA (left) and the 10,000 year return levels (right) estimated using this approach \citep{winter2017rfa}.}
    \label{fig:RFA}
\end{figure}

The main drawback of RFA is the fixed specification of regions deemed to be statistically homogeneous. Cells within each homogeneous region are pooled together for parameter estimation. However, this process creates artificial boundaries, meaning that the cells along these boundaries can potentially have very different characteristics from neighbouring cells that have been assigned to other regions. Unless there is a physical analogue of these statistical boundaries, for example, a mountain range, then the intuition behind this specification of spatial similarity begins to break down. Figure~\ref{fig:RFA} (left) shows how the classification algorithm of \citet{weiss2014formation} defines the homogeneous regions over our spatial domain. \\
\vspace{-10pt}

RFA assumes a constant shape parameter over all cells in each homogeneous region, which is an arguably simplistic assumption to make. When cell-level parameter estimates are used to determine return levels, spatially inconsistent estimates are produced (see Figure~\ref{fig:RFA} (right)). The large return levels obtained for the south-east region are unrealistic as they are less likely to experience precipitation at the same level as Scotland on a regular basis. This anomaly is most likely caused by the influence of the maxima caused by convective events, as observed in Figure~\ref{fig:exploratory}. When the data over the south-east region are pooled together, these events are determining the heaviness of the tail over the entire region under the assumption of a common shape parameter. \\

This approach brings about substantial reductions in the statistical estimate of return level uncertainty. However, this uncertainty has not been adjusted to account for the misspecification of the model, which comes from specification of the homogeneity of the regions and independence between all sites within a homogeneous region. Failure to account for this misspecification results in misleadingly narrow estimated return level confidence intervals, and perhaps inaccurate design specifications for practitioners as a consequence.



\section{Model}
\label{sec:model}
Our approach uses a Bayesian spatial hierarchical model to induce similarity of parameter estimates of an extreme value model between neighbouring sites. The model consists of a data, process and prior level.
\subsection{Data level}
\label{subsec:data_level}
On the top layer of the hierarchy, we fit a threshold excess model to the tails of the precipitation distribution in each cell.
Let $Y_{j,t}$ be the daily precipitation level in cell $j \in \{1,\hdots,d)$ on day $t$. We assume that the excesses above a threshold $u_j$ follow a GPD distribution. The threshold $u_j$ is typically chosen using standard selection diagnostics \citep{coles2001introduction} or by choosing an appropriately large quantile. We can thus write
\[ Y_{j,t} \mid\left( Y_{j,t} > u_j, \tilde{\sigma}_j, \xi_j\right) \sim \text{GPD}( \tilde{\sigma}_j, \xi_j),\]
where $\tilde{\sigma}_j = {\sigma_u}_j - \xi_j u_j$ denotes the threshold-independent scale parameter and $\xi_j$ denotes the shape parameter in cell $j$. We assume that $Y_{j,t} \mid Y_{j,t} > u_j$ given $( \tilde{\sigma}_j, \xi_j)$ is conditionally independent of $Y_{i,t} \mid Y_{i,t} > u_i$ for all $i \neq j$ and all $t$. While this conditional independence assumption is common in hierarchical modelling, it is often not well-supported in precipitation-based applications as storms can affect multiple locations simultaneously, for example. We proceed with the assumption as our interest lies in the marginal extremal characteristics in each cell. However, we include a magnitude adjustment in the GPD likelihood to account for the misspecification of conditional independence, which allows for appropriate inference \citep{ribatet2012bayesian}. To be precise, we define the adjusted likelihood $L^{*}(\boldsymbol{\theta},\boldsymbol{y})$ such that
\begin{equation}
  L^{*}(\boldsymbol{\theta},\boldsymbol{y}) =  {L(\boldsymbol{\theta},\boldsymbol{y})}^{k},
  \label{eq:adjlik}
\end{equation}
where $k>0$ and $L(\boldsymbol{\theta},\boldsymbol{y}) = \prod_{j=1}^{d} L(\boldsymbol{\theta}_j,\boldsymbol{y}_j)$, the product of GPD likelihoods over all sites. This represents a magnitude adjustment of the likelihood that leaves parameter estimates unchanged, but scales the uncertainty of these estimates to account for model misspecification. \\

The adjustment is based on the property that $D_{full} \rightarrow \chi_{p}^{2}$ as $n \rightarrow \infty$, where $D_{full}$ denotes the deviance function corresponding to the full likelihood, $p$ is the number of parameters and $n$ is the number of data points. For the deviance $D_{mis}$ corresponding to the misspecified likelihood~(\ref{eq:adjlik}), $D_{mis} \rightarrow k \sum_{i=1}^{p} \lambda_i X_i$ as $n \rightarrow \infty$, where $X_1, \hdots, X_p$ are independent $\chi_{1}^{2}$ random variables and $\lambda_1, \hdots, \lambda_p$ are the eigenvalues of the Godambe information matrix. Setting $k=p/\sum_{i=1}^{p} \lambda_i$ ensures that $\mathrm{E}[D_{mis}]$ converges to $\mathrm{E}[\chi_{p}^{2}] = p$. However, in general, higher moments do not match those of $\chi_{p}^{2}$. \citet{ribatet2012bayesian} propose a curvature adjustment to ensure full convergence to a $\chi^2$ distribution. However, the magnitude adjustment is simpler to implement and performs well in practice so this is the approach we will take. Estimation of $k$ requires estimation of the Godambe matrix, the details of which can be found in \citet{ribatet2012bayesian} and \citet{varin2011overview}. The quantity $k$ can be interpreted as the effective proportion of locations that have independent data. Due to spatial dependence, we expect $k$ to be in the interval $(1/d,1]$, where a value of $k=1$ corresponds to the data being iid over space, and where $k=1/d$ corresponds to perfect dependence, with effectively one location's worth of information in the dataset. \\

We also make a working assumption that the precipitation data are iid over observations in each cell. Partial autocorrelation plots show that these data exhibit signs of a third-order temporal dependence structure, meaning that the level of precipitation on a given day will be influenced by levels on the previous three days. Declustering methods are typically used to identify independent clusters of precipitation events, see \citet{ferro2003inference} for details of such a method. However, as our focus is on modelling spatial dependence, we assume that the data in each cell are iid in time.

\subsection{Process level}
This layer of the hierarchical model induces the borrowing of strength across locations. We assume an underlying spatial process in the mean of the distribution of both GPD parameters, such that the parameters in a cell are more likely to be similar to those in neighbouring cells than those further away. As well as these spatial effects, we can also incorporate fixed climate or physical effects in a cell through the inclusion of covariates in the model. Formally, we assume
\[ \boldsymbol{\theta}_j \sim N(\boldsymbol{X}_j \boldsymbol{\beta} + \boldsymbol{\phi}_j, T_{\boldsymbol{\theta}}^{-1}), \] 
where $\boldsymbol{\theta}_j = (\log \tilde{\sigma}_j,\xi_j)$ is the vector of transformed GPD parameters in cell $j$, $\boldsymbol{X}_j$ is the design matrix of fixed covariates in cell $j$, $\boldsymbol{\beta}$ is the vector of regression coefficients, $\boldsymbol{\phi}_j$ is a spatial random effect in cell $j$ and $T_{\boldsymbol{\theta}}$ is the common precision matrix for the GPD parameters over all cells. We include a log-link in the estimation of the scale parameter to ensure positivity in the presence of covariate information. \\

The latent spatial process is induced through fixed effect terms $\boldsymbol{X}_j$ and the random effect term $\boldsymbol{\phi}_j$. Since we are working with lattice data, the natural choice of prior for $\boldsymbol{\phi}_j$ is a multivariate conditional autoregressive (CAR) model. We make the assumption that the lattice structure of the region represents a Markov random field, and that specification of local conditional relationships can be used to specify a global joint distribution \citep{banerjee2014hierarchical}. The Intrinsic Autoregressive (IAR) model is a special case of the CAR model, with the conditional prior for $\boldsymbol{\phi}_j$ written as
\[\boldsymbol{\phi}_j \mid \boldsymbol{\phi}_{(-j)} \sim N\left(\frac{1}{m_j}\sum_{i \in \delta_j} \boldsymbol{\phi}_i, \frac{1}{m_j} {T_{\boldsymbol{\phi}}}^{-1}\right), \]
where $\delta_j$ denotes the set of neighbours of cell $j$, $m_j$ is the number of neighbours and $T_{\boldsymbol{\phi}}$ denotes the common precision matrix of random effects over all cells. The IAR model has an improper density and does not represent a legitimate probability distribution. For this reason, it cannot be applied directly to data and is often used as a prior, since inference can still be made as long as the posterior distribution is proper. With IAR priors, the fixed effects are identifiable as long as the spatial random effects are centred so that $\sum_{j=1}^{d} \boldsymbol{\phi}_j = 0$ \citep{banerjee2014hierarchical}. This can be implemented in practice by centering each joint iteration of the MCMC sampling of the random effect terms. An alternative to the IAR approach is to specify a propriety parameter $\rho$, which controls the level of spatial association between cells. \citet{cooley2010spatial} use a separable formulation to specify the joint prior density of $\boldsymbol{\phi} = (\boldsymbol{\phi}_1, \hdots, \boldsymbol{\phi}_d)$ using matrix decomposition techniques from \citet{rue2005gaussian}.

\subsection{Prior level}
With the exception of the random effect parameters inducing spatial dependence, we have no prior knowledge regarding any of the parameters in the model. We choose to assign uninformative priors where possible. We also choose to assign conjugate priors where possible to allow us to sample from the posterior distribution using Gibbs sampling. For the regression coefficients, we use an empirical Bayes approach, such that
\[ \boldsymbol{\beta} \sim N(\boldsymbol{\beta}_0, T_{\boldsymbol{\beta}}^{-1}), \]
where $\boldsymbol{\beta}_0$ and $T_{\boldsymbol{\beta}}^{-1}$ denote the prior mean and precision matrix of the regression coefficients respectively. The intercept terms are taken to be the means of the cell-wise maximum likelihood estimates for $\log \tilde{\boldsymbol{\sigma}}$ and $\boldsymbol{\xi}$ while the covariate coefficients are given a mean of 0. The precision matrix is chosen so that the intercept terms have a precision of 0.01 and the covariate effects have a precision of 0.1. These values are typically chosen to represent the levels of variability one would expect in the parameter estimates. Because we cannot achieve conjugacy for the GPD parameters, we impose a flat joint prior such that $\pi(\boldsymbol{\theta}_j) \propto 1/\sigma_j$. \\

We assign conjugate Inverse-Wishart priors to both ${T_{\boldsymbol{\theta}}}^{-1}$ and ${T_{\boldsymbol{\phi}}}^{-1}$, such that:
\begin{eqnarray*}
{T_{\boldsymbol{\theta}}}^{-1} & \sim & \text{Inv-Wishart}\left(\nu_\theta,\Omega_\theta\right) \\
{T_{\boldsymbol{\phi}}}^{-1} & \sim & \text{Inv-Wishart}\left(\nu_\phi,\Omega_\phi\right).
\end{eqnarray*}
Matrices $\Omega_\theta$ and $\Omega_\phi$ were chosen to reflect the levels of variability found within each cell's parameter estimates and also between each cell's parameter estimates. 
\subsection{Implementation}
As we have used conjugate priors for some parameters, we can construct closed-form full conditionals from these and thus construct these components of the joint posterior distribution using a Gibbs sampler. The full conditionals are
\begin{align*}
\boldsymbol{\beta} \mid \hdots &\sim  N\left(\boldsymbol{\mu}_\beta,\boldsymbol{V}_\beta\right) \\
\boldsymbol{V}_\beta &=  {\left[T_{\boldsymbol{\beta}} + \boldsymbol{X}^{T} T_{\boldsymbol{\theta}} \boldsymbol{X} \right]}^{-1} \\
\boldsymbol{\mu}_\beta &=  \boldsymbol{V}_{\beta} \left[T_{\boldsymbol{\beta}} \boldsymbol{\beta}_0 + \boldsymbol{X}^{T} T_{\boldsymbol{\theta}} (\boldsymbol{\theta} - \boldsymbol{\phi})\right] \\[10pt]
\boldsymbol{\phi}_j \mid \hdots &\sim  N\left(\boldsymbol{\mu}_{\phi}, \boldsymbol{V}_{\phi}\right) \\
\boldsymbol{V}_{\phi} &= {\left[m_j T_{\boldsymbol{\phi}} + T_{\boldsymbol{\theta}}\right]}^{-1} \\
\boldsymbol{\mu}_{\phi} &= \boldsymbol{V}_{\phi} \left[ T_{\boldsymbol{\theta}} (\boldsymbol{\theta}_j - \boldsymbol{X}_j \boldsymbol{\beta}) + T_{\boldsymbol{\phi}} \sum_{i \in \delta_j} \boldsymbol{\phi}_i \right] \\[10pt]
{T_{\boldsymbol{\theta}}}^{-1} \mid \hdots &\sim  \text{Inv-Wishart}\left(\nu_\theta + d, \Omega_\theta + {(\boldsymbol{\theta} - \boldsymbol{\phi} - \boldsymbol{X} \boldsymbol{\beta})}^{T} {(\boldsymbol{\theta} - \boldsymbol{\phi} - \boldsymbol{X} \boldsymbol{\beta})} \right) \\[10pt]
{T_{\boldsymbol{\phi}}}^{-1} \mid \hdots &\sim \text{Inv-Wishart} \left(\nu_\phi + d, \Omega_\phi + \boldsymbol{\phi}^{T} \boldsymbol{W} \boldsymbol{\phi} \right), 
\end{align*}
where $\boldsymbol{W}$ is a matrix defining spatial proximity between cells. The matrix $\boldsymbol{W}$ has off-diagonal elements $w_{ij} = -1$ if cells $i$ and $j$ are adjacent and $w_{ij}=0$ otherwise and diagonal elements $w_{ii}= - \sum_{i \neq j} w_{ij}$, being the number of neighbours of cell $i$. \\

Proposals for $\boldsymbol{\beta}$, $\boldsymbol{\phi}_j$, ${T_{\boldsymbol{\theta}}}^{-1}$ and ${T_{\boldsymbol{\phi}}}^{-1}$ can be generated using a Gibbs sampler step. The GPD parameters are constructed using a Metropolis-Hastings step. We use a random walk Metropolis scheme with a Gaussian proposal distribution, tuning the proposal covariance matrix appropriately to give the optimal acceptance rate of \citet{roberts2001optimal}.

\section{Results}
\label{sec:results}

In this section, we compare output from the Bayesian hierarchical modelling approach with that of analysing each cell separately, to assess whether there is any value in including a latent spatial process in the modelling procedure. We select the $95\%$ quantile as our extreme value threshold for each cell. Threshold stability plots were checked at a handful of cells and the aforementioned threshold represented a sensible choice in all cases. For the single cell analysis, we impose an uninformative Uniform prior on the GPD parameters. Previous studies on extreme rainfall have used the prior density defined by \citet{martins2000generalized}, which is useful in small-sample cases as it constrains the shape parameter to be in a sensible interval. In this analysis, however, we obtained realistic shape parameters for estimating rainfall, so we chose not to use this prior. To account for the misspecification of conditional independence between cells, we use the adjusted likelihood~(\ref{eq:adjlik}) in the hierarchical model. The constant $k=0.801$ is estimated using methods described in Section~\ref{subsec:data_level} with the adjusted likelihood being used to form the posterior. This value of $k$ corresponds to a relatively weak level of spatial dependence within the data, with effectively approximately 145 locations worth of independent information in the 181 cells. In both models, we run the MCMC for 20,000 iterations and discard the first 5,000 as burn-in. Parameter estimates in this analysis are taken to be posterior means. \\
\begin{figure}[h!]
    \centering
    \includegraphics[width=12cm]{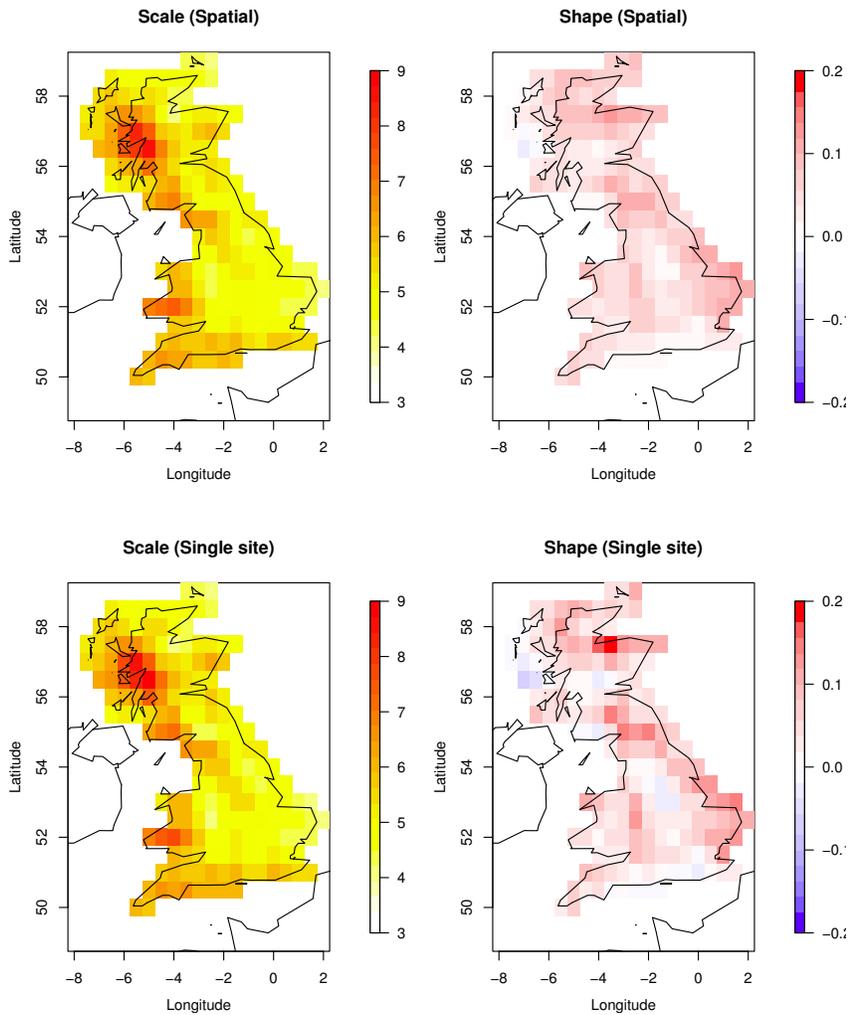}
    \caption{Parameter estimates of the threshold-dependent scale and shape parameters for all cells from the Bayesian hierarchical model (top) and analysing each cell separately (bottom).}
    \label{fig:spat_SS_par}
\end{figure}

As we are working on a relatively coarse lattice, there is a limited amount of physical information we can incorporate into the problem without being overly simplistic regarding the assumptions we make. For example, including elevation as a covariate would be unwise because various different landscapes can be observed in a single cell, and assigning an average to that would mask out any meaningful effect. We assessed whether adding latitude and longitude as covariates  would improve the overall fit using the deviance information criterion (DIC) \citep{spiegelhalter2002bayesian}.  However, including these covariates had no significant effect. It is possible that the study region is too small for any substantial latitude/longitude influence to be significant. We also checked if grid cell distances from both west and east coasts improved the model fit but the null model was preferred. The results that follow are from the null model in the absence of any fixed covariate information. \\

Figure~\ref{fig:spat_SS_par} compares the GPD parameter estimates from the hierarchical model and from analysing each cell separately. In both models, the scale parameter reflects the claim from the exploratory analysis that larger precipitation events tend to occur on the west coast of Britain, with particular impact in Scotland. The heavy tails in the south east of England are likely determined by short-term, localised convective events. While there is a little visible difference in estimation of the scale parameter, the hierarchical model produces a much smoother surface for the shape parameter. This parameter has a clear geographical structure, with higher values in the south-east of England and smaller values on the south coast and parts of the Midlands, implying that an assumption of constancy is overly simplistic. The shape parameter is typically difficult to estimate given short data records and benefits from the extra information supplied through the spatial prior. \\


The true scientific value of this study is only clear when analysing return levels rather than the GPD parameters themselves. By rearranging the survivor function and setting equal to $r^{-1}$ such that
\begin{equation}
    n_y \lambda_u {\left(1+\frac{\xi x}{\sigma_{u}}\right)}^{-1/\xi} = r^{-1},
\end{equation}
where $n_y$ denotes the number of observations in a given year, we solve for $x$ to obtain the $r$-year return level, that is, the value that is exceeded on average once every $r$ years. The posterior distribution of the $r$-year return level is obtained by applying this function to every MCMC iteration (after burn-in) of the GPD parameter chains. This enables us to extract posterior confidence intervals in a natural way by looking at the quantiles of the estimated posterior. \\

\begin{figure}[H]
    \centering
    \includegraphics[width=10.5cm]{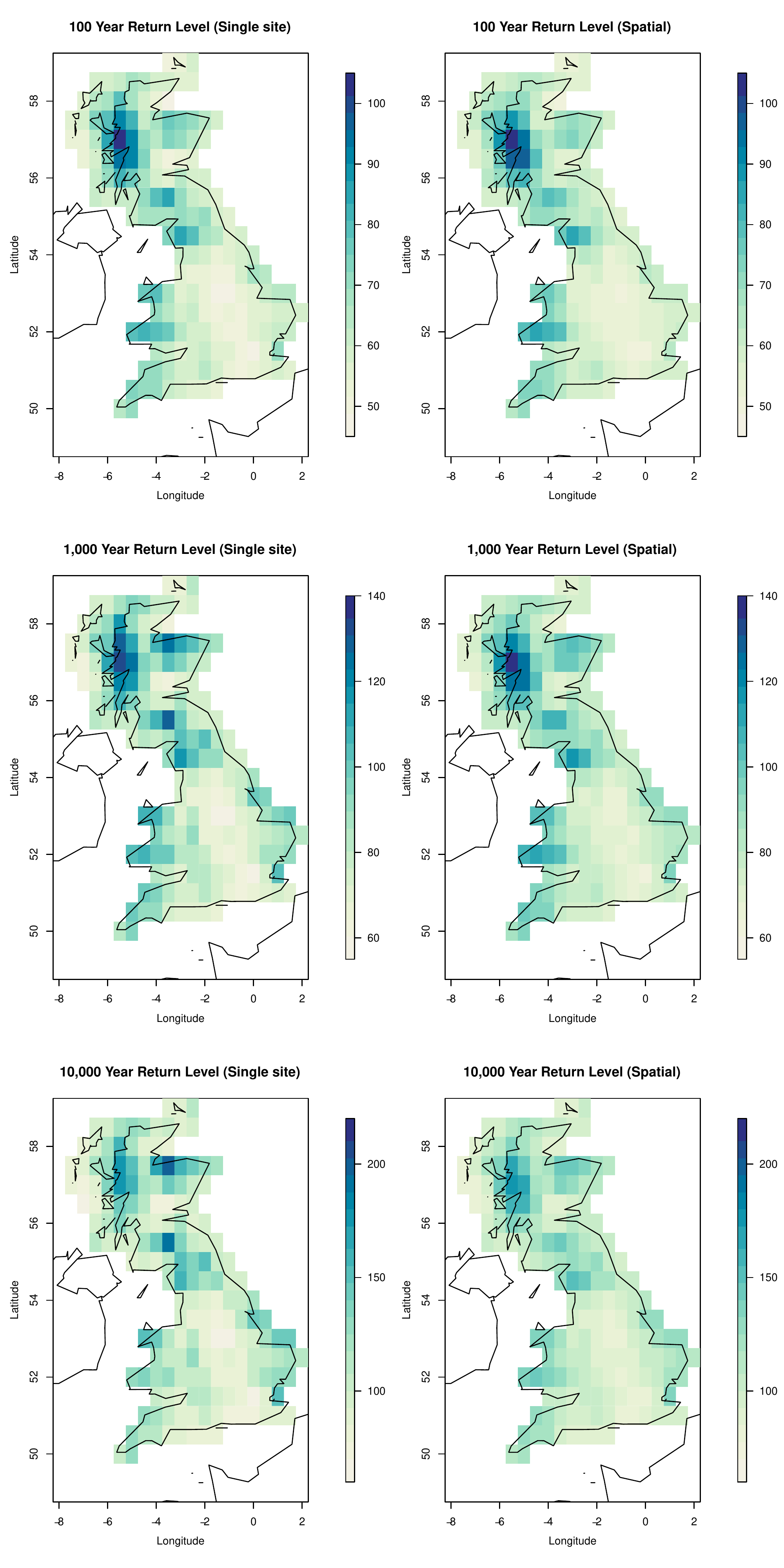}
    \caption{\vspace{-10pt} 100- (top), 1,000- (middle) and 10,000- (bottom) year return level estimates for each cell under the single cell analysis (left) and hierarchical model (right).}
    \label{fig:ret_lev}
\end{figure}

The $r$-year return level is a more intuitive quantity for practitioners to analyse, as these estimates can be considered in the design of infrastructure to defend against extreme precipitation events. Figure~\ref{fig:ret_lev} shows that the spatial model produces similar return level estimates to that of the single cell analysis when the return period is short. As we move far beyond the range of the data, the dominance of the shape parameter means that we are more likely to see spatially inconsistent return level estimates when performing a single cell analysis. The spatial model performs well in the sense that we move further into the tail in a smooth and realistic way, which is also spatially coherent. \\

Both RFA and the hierarchical modelling approach pool information across neighbouring locations to improve spatial consistency of parameter estimates and reduce uncertainty of these estimates. The methods differ in the imposition of the spatial process on the GPD parameters. The artificial boundaries arising from the specification of homogeneous regions in RFA can result in spatially inconsistent return level estimates (see Figure~\ref{fig:RFA}). The assumption of a constant shape parameter in each region means that the most extreme events can determine the size of the tail over the entire region, which may not be realistic. The hierarchical modelling approach, under the weaker and more intuitive assumption of similarity across neighbouring locations, produces a more realistic return level landscape, representing a smoother extrapolation into the tails of the distribution of precipitation. \\

\begin{figure}[h!]
    \centering
    \includegraphics[width=14cm]{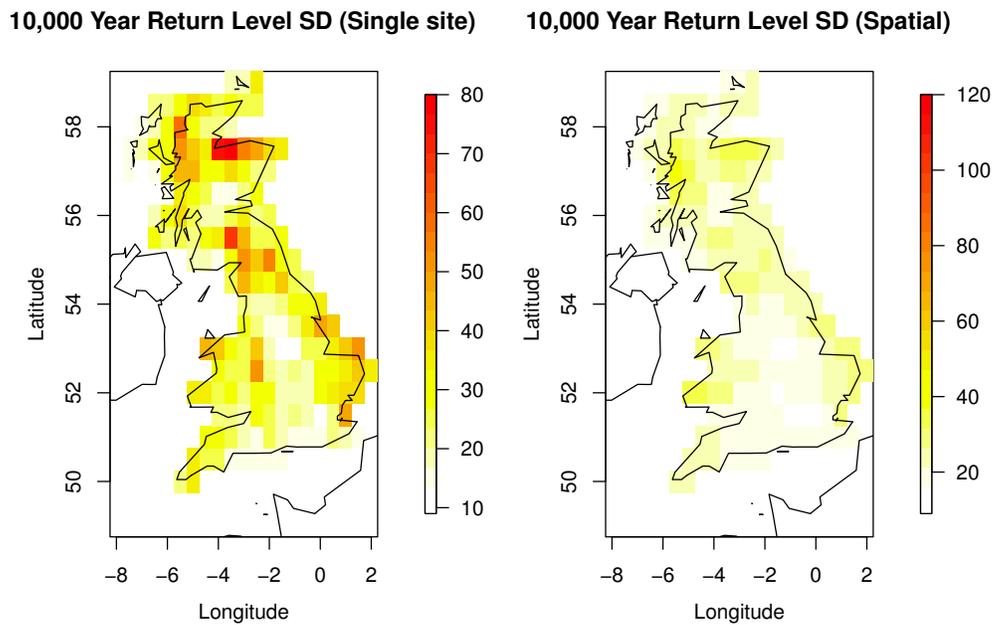}
    \caption{Standard deviations of the posterior distribution of the 10,000-year return level for the single cell analysis (left) and hierarchical model (right).}
    \label{fig:uncertainty}
\end{figure}

As well as improving spatial consistency, spatial hierarchical models borrow strength across locations, reducing uncertainty in return level estimates. Figure~\ref{fig:uncertainty} shows the standard deviation of the estimated posterior 10,000-year return level distribution. In all cases, this measure is reduced by using the spatial model, and in some cases, substantially. The largest standard errors tend to be observed at coastal locations, which is intuitively a consequence of the increased prior variance caused by a smaller number of neighbours compared to inland locations. The uncertainty in these estimates is quantified correctly under the misspecification of the model using the adjusted likelihood~(\ref{eq:adjlik}). Without this adjustment, the estimated confidence intervals of return levels are narrow and can result in misleading inference.  \\

\begin{figure}[h!]
    \centering
    \includegraphics[width=7cm]{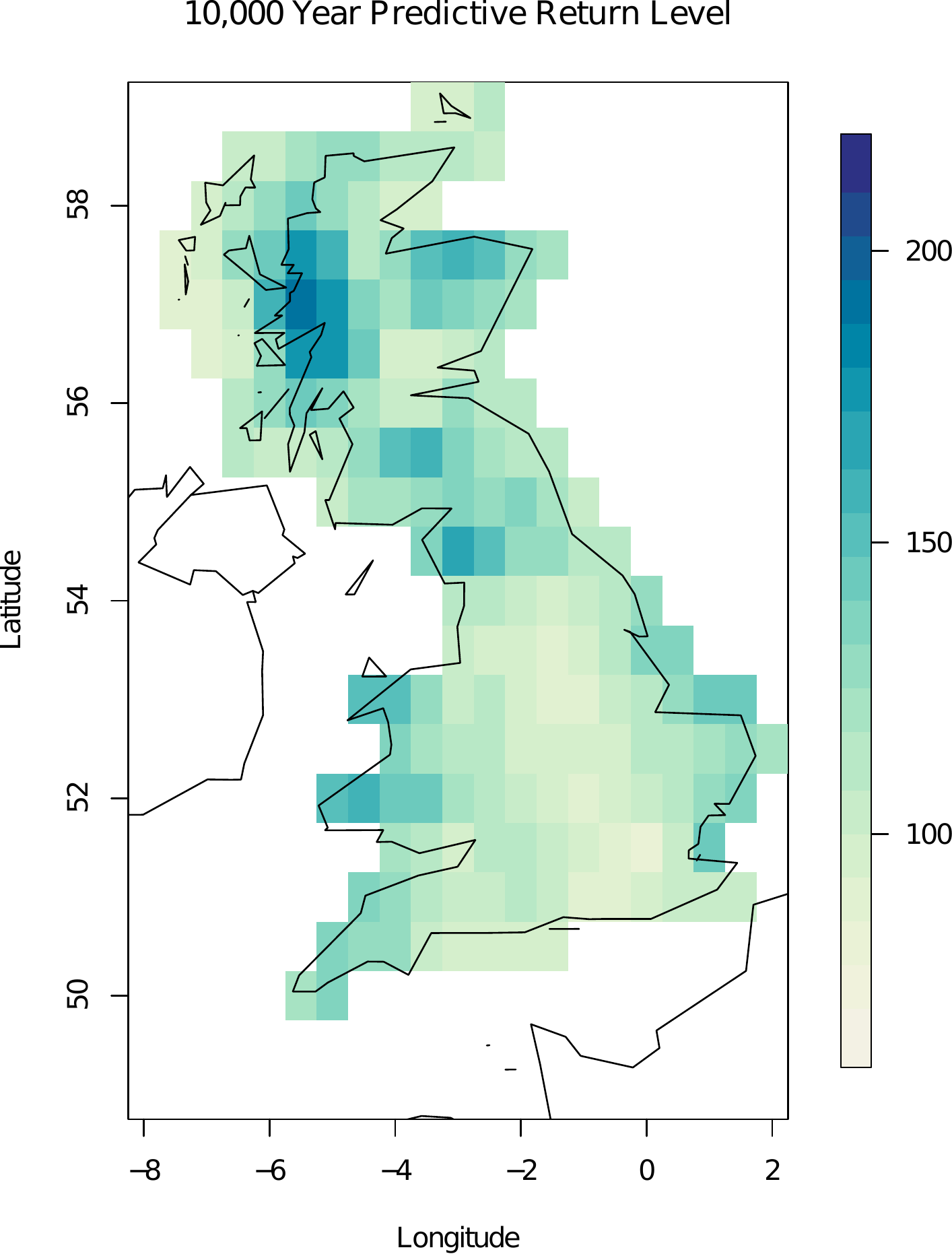}
    \caption{The 10,000-year predictive return level for all cells.}
    \label{fig:predictive}
\end{figure}

The Bayesian paradigm allows us to handle the issue of prediction in a natural way. We can construct the distribution of a future threshold excess - the \emph{predictive distribution}. This incorporates both parameter uncertainty and randomness in future observations. For cell $j$, we have that
\[ \Pr(Y_j \leq \tilde{y}_j \mid \boldsymbol{y}) = \int_{\boldsymbol{\theta}_j} \Pr(Y_j \leq \tilde{y}_j \mid \boldsymbol{\theta}_j) \pi(\boldsymbol{\theta}_j \mid \boldsymbol{y}) \mathrm{d}\boldsymbol{\theta}_j. \]
We can then define the $r$-year predictive return level to be the value of $\tilde{y}_j$ that satisfies
\[ \Pr(Y_j \leq \tilde{y}_j \mid \boldsymbol{y}) = 1- r^{-1}. \]
While this is analytically intractible, we can approximate using a Monte Carlo summation of the samples from the estimated posterior distribution. This gives
\[ \Pr(Y_j \leq \tilde{y}_j \mid \boldsymbol{y}) \approx \frac{1}{N} \sum_{i=1}^{N} 
\Pr(Y_j \leq \tilde{y}_j \mid \boldsymbol{\theta}_{j}^{(i)}), \]
where $N$ denotes the number of MCMC samples after the burn-in has been discarded. The predictive return level estimate can then be evaluated using a standard numerical solver \citep{tawn1996bayesian}. The $10,000$-year predictive return level for all cells is shown in Figure~\ref{fig:predictive}. The values have the same spatial pattern as that of the mean of the posterior distribution of $10,000$-year return levels, but the magnitude of the values is slightly higher. This is due to the additional parameter uncertainty being accounted for in the predictive analysis.

\section{Discussion}
\label{sec:disc} 
We have developed a flexible model that estimates precipitation return levels at multiple locations in a spatially consistent manner. Inducing spatial similarity through a prior model allows us to borrow strength across locations and reduce uncertainty in return level estimates. The prior model helps to build a smooth map of return level estimates, without the need to impose boundary restrictions required for regional frequency analysis. \\

Since the model is flexible by nature, there are many improvements one could make to this analysis. If one was looking at cells on a finer grid, one could include more physical covariates related to the geography of the region. Previous studies have shown that elevation is a key influence on the size of a rainfall event \citep{cooley2010spatial}. Another extension of this work could be the inclusion of a seasonal component in the model. It is widely known that heavy rainfall events in the southeast are largely the result of short-term, localised convective storms that occur in summer. In contrast, heavy rainfall events in northern Scotland, for example, tend to occur as part of a more sustained, synoptic storm system from the Atlantic Ocean. It would be useful to be able to discern between these two systems in the model. This involves introducing a temporally varying covariate, adding an extra dimension to the model. \citet{economou2014spatio} models sea-level pressure using both spatially and temporally varying covariates in a Bayesian hierarchical model framework. Seasonality can been modelled as a periodic covariate using Fourier series \citep{jonathan2011modeling} in the extreme value parameters. Alternatively, \citet{fawcett2006hierarchical} models extreme precipitation for multiple sites on a month-by-month basis but induces some temporal smoothness through a seasonal random effect term modelled through a conditional autoregressive prior in time. \\

It is also possible to consider further development of the spatial structure. Here, we assume a first order neighbourhood structure is sufficient in summarising the spatial dependence inherent in the parameters. \citet{jalbert2017spatiotemporal} proposes using a second-order Markov random field, which gives smoother results than the first-order structure. However, their assessment was over a wider spatial domain and thus their assumption of a higher-order structure was justified. \\

A drawback of previous studies using the Bayesian hierarchical framework concerned the inference arising from the misspecified conditional independence likelihood. However, our approach allows appropriate inference and uncertainty quantification to be carried out, avoiding unrealistically narrow confidence intervals. Advancing beyond the assumption of conditional independence, recent studies have included a max-stable process in the data layer of a hierachical model \citep{ribatet2012bayesian,reich2012hierarchical,thibaud2016bayesian}.
This approach captures the local spatial dependence of the extremes, and provides a mechanism for simulating realistic fields of precipitation over space.
\section*{Acknowledgements}
The authors extend their thanks to Jonathan Tawn for very helpful comments and support. We gratefully acknowledge the support of the EPSRC funded EP/H023151/1
STOR-i Centre for Doctoral Training, EDF Energy and the Met Office. Paul Sharkey was partially supported by the Joint UK BEIS/Defra Met Office Hadley Centre Climate Programme (GA01101).

\bibliographystyle{apa}      
\bibliography{refs} 

\end{document}